\documentclass[aps,pra,twocolumn,showpacs]{revtex4}
\usepackage{graphicx}
\usepackage{bm}
\usepackage{amsfonts}
\usepackage{amssymb}
\usepackage{amsmath}
\usepackage{times}
\usepackage{latexsym}
\usepackage{amsbsy}
\usepackage{pifont}
\usepackage{computer2}

\begin{document}
\setlength\arraycolsep{1pt}

\title{Entangled Photons from Small Quantum Dots}

\author{P.M.\ Visser, K.\ Allaart and D.\ Lenstra}
\affiliation{Vrije Universiteit Amsterdam, De Boelelaan 1081, 1081HV Amsterdam, The Netherlands}
\email{PMV@nat.vu.nl,allaart@nat.vu.nl,lenstra@nat.vu.nl}

\begin{abstract}
We discuss level schemes of small quantum-dot turnstiles and their applicability in the production of entanglement in two-photon emission. Due to the large energy splitting of the single-electron levels, only one single electron level and one single hole level can be made resonant with the levels in the conduction band and valence band. This results in a model with nine distinct levels, which are split by the Coulomb interactions. We show that the optical selection rules are different for flat and tall cylindrically symmetric dots, and how this affects the quality of the entanglement generated in the decay of the biexciton state. The effect of charge carrier tunneling and of a resonant cavity is included in the model.
\end{abstract}
\pacs{42.50.Ct,42.55.Sa,73.21.La,85.35.Gv.}

\maketitle

\section{Introduction}

The constant progress in the fabrication of nanostructures has led to novel semiconductor devices, like quantum wires and quantum dots, that allow the confinement and control of single electrons. In quantum dots it is possible to experimentally control the tunneling of single electrons and holes. These systems exhibit quantum-correlations in the emission statistics \cite{Becher,Ehrenfreund}, and are very promising for future applications in quantum communication.

A quantum dot that emits single photons controlled by the switching of a voltage is called a single-photon turnstile \cite{Yamamoto1}. In such a system, the quantum dot is allowed to contain at most one single electron-hole pair, so that one photon is created at a time. In order to realize this, one makes use of the Coulomb blockade effect to suppress tunneling of a second electron or hole onto the dot. This implies that the system must be cooled to temperatures with $k_{\rm B} T$ smaller than the Coulomb splittings. In a two-photon turnstile, two electron-hole pairs are created, before two successive photons are emitted. Recently, a two-photon turnstile has been proposed \cite{Yamamoto2} as a device to generate entangled photon pairs, which makes these systems very interesting. Because the Pauli principle allows occupation of an electronic level by at most two electrons, a two-photon turnstile can be realized without Coulomb blockade effects, provided that the thermal energy is smaller than the splitting of the single-particle levels so that tunneling of more than two electron-hole pairs is avoided. In a small quantum dot, this splitting can be much larger then the Coulomb splittings, so that a two photon turnstile does not require cooling in the milliKelvin regime for proper operation.

In order to explore various possibilities for generation of entangled photons by two-photon turnstiles, we consider in this paper simple level schemes that can occur when a quantum dot is smaller than the bulk exciton size, and study which situations are favorable for the generation of entangled photon pairs. In this regime, the electron and hole wave functions are strongly localized so that the energy separation of individual levels is larger than the characteristic Coulomb interaction energies. The central idea is that, due to resonant tunneling, only one twofold degenerate electron level and one twofold degenerate hole level of the quantum dot play an active role. Other electron and hole levels are too remote in energy and may therefore be discarded in a first approximation. Within the resulting finite scheme of dot states we study the entanglement of cascade photons from the biexciton decay, notably its dependence on the competition between charge carrier tunneling and radiative electron-hole recombination. The effect of a resonant cavity will also be calculated.

\section{Models for Small Quantum Dots}

\begin{figure}[htb]
\centerline{\includegraphics[width=6cm]{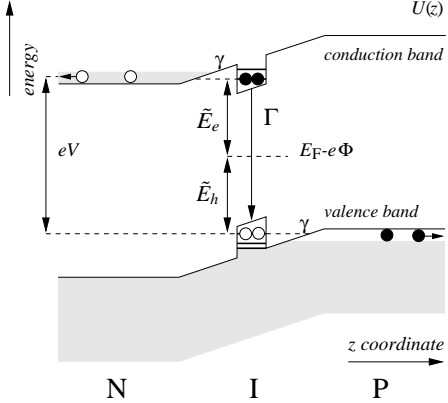}}
\caption[]{Energy-band structure of the PIN junction, for a cross section through the quantum dot along the $z$ axis. The quantum dot is a small cylindrical structure located in the I layer between N and P semiconductors. With the gate potential $\Phi$, the electron and hole energies can be shifted. A bias voltage $V$ over the junction allows electrons (black) and holes (white) to tunnel across the barriers with rate $\gamma$ and $\Gamma$ is the photon emission rate, as indicated.}
\label{Fig1}
\end{figure}

A diagram of the semiconductor structure that we have in mind is shown in Figure \ref{Fig1}. The quantum dot is located between P doped and N doped material. By means of a bias voltage $V$ over the junction and a gate voltage $\Phi$ of an electrode near the dot, the electron level of the dot is made resonant with the bottom of the conduction band of the N type material and the hole level of the dot with the top of the valence band of the P type material. The quasi-particle energies $\tilde E_e$ and $\tilde E_h$ are well defined as the energy of the dot with one excess electron, respectively hole, with respect to the neutral state. The bias voltage $V$ separates the energy levels between the N and P sides by $eV$ and the gate voltage shifts the electron and hole levels by $-e\Phi$ and $e\Phi$. The resonance condition for a cental dot is then
\begin{equation}
\tilde E_e - e\Phi = eV/2 , \;\; \tilde E_h + e\Phi = eV/2 .
\label{4}
\end{equation}
It is energetically favorable that electrons tunnel into the dot when $eV/2>\tilde E_e-e\Phi$ and out of the dot when $eV/2<\tilde E_e-e\Phi$. The resulting level scheme is shown in Figure \ref{Fig2} and has only sixteen basis states, part of which are charged due to the presence of one or two excess electrons or holes. In this scheme, the state with highest energy is the bi-exciton state, with two electrons in the upper level and two holes in the lower level. The optical properties are determined by exciton and biexciton states \cite{Bayer,Abram}. It is in the cascade decay from the biexciton to the ground state via a state of the one-exciton multiplet that an entangled photon pair may be generated. The splitting of the one-exciton multiplet is an effect of the Coulomb interaction between the particles \cite{Kouwenhoven}, which will be discussed in the following. 

When one switches the gate voltage $\Phi$ or the bias voltage $V$, during a short time interval, first to a higher value and immediately thereafter to a lower value, then one promotes the tunneling of electrons from the N type material into the upper level of the dot, immediately followed by tunneling of holes from the P type material \cite{Yamamoto3}. The system state then follows the path indicated by the diagonal arrows in Fig.\ \ref{Fig2}. The bi-exciton state is produced without intermediate formation of a one-exciton state. Ideally the system will therefore emit a cascade of two photons, one on transition $1$ and one on transition $2$, with frequencies $\omega_1$ and $\omega_2$. It is important to switch the gate voltage $\Phi$ (and bias voltage) back to the resonant values (\ref{4}) immediately after the preparation of the biexciton, in order to reduce the probability of electron or hole tunneling before the second photon is emitted. If tunneling nevertheless occurs before the second photon is emitted, then the transitions $3$ or $4$ between the charged states may occur. Their frequencies $\omega_3$ and $\omega_4$ are in general different from $\omega_1$ and $\omega_2$, as may be checked from Eq.\ (\ref{1a}) for the energy levels. Therefore these photons may in principle be filtered out. In the next section we shall solve a master equation which includes the competition between tunneling and recombination processes. For small dots, states with more electrons or holes have non-resonant energies. Provided the bias voltage is not too large, the biexciton state is the highest excited state and no multi-exciton states are formed \cite{Hawrylak}.

\subsection{Level Splitting and Photon Energies}

We consider the Coulomb interaction between electrons and holes as a perturbation on interaction-free levels. This approximation is well known in the context of quantum dots without holes in the electronic distribution, which are in (near) equilibrium \cite{Beenakker,Koch}. For semiconductor quantum dots with holes, however, localized exciton states are formed from electron-hole pairs. Only if the dot is smaller than the exciton size the single particle levels are well defined and the Coulomb interaction can be treated as a perturbation \cite{Koch,Zunger}. This is the condition that we suppose to be fulfilled in the following. Since we consider systems in absence of magnetic fields, the single-electron levels are twofold degenerate due to time-reversal symmetry \cite{Kittel}. Let the two degenerate single-particle states for the electron and hole level be described by
\begin{eqnarray}
|e\rangle &=& \int\!\!d\vec r\, |\vec r\rangle \Big( |\uparrow\rangle\psi_1(\vec r) + |\downarrow\rangle\psi_2^*(\vec r) \Big) ,
\nonumber \\
|\bar e\rangle &=& \int\!\!d\vec r\, |\vec r\rangle \Big( |\uparrow\rangle\psi_2(\vec r) - |\downarrow\rangle\psi_1^*(\vec r) \Big) ,
\nonumber \\
|h\rangle &=& \int\!\!d\vec r\, |\vec r\rangle \Big( |\uparrow\rangle\chi_1(\vec r) + |\downarrow\rangle\chi_2^*(\vec r) \Big) ,
\nonumber \\
|\bar h\rangle &=& \int\!\!d\vec r\, |\vec r\rangle \Big( |\uparrow\rangle\chi_2(\vec r) - |\downarrow\rangle\chi_1^*(\vec r) \Big) ,
\label{2}
\end{eqnarray}
in terms of the wave functions $\psi_j(\vec r)$ and $\chi_j(\vec r)$ for the spinor components. Due to spin-orbit coupling these single particle states are not spin eigenstates in general. The dot states that form the basis of the configuration space are then described by the occupation of the four basis states (\ref{2}). The number of electrons within this space thus ranges from zero up to four. The ground state has the hole level occupied with electrons and therefore is an effective quasi-particle vacuum, denoted with $|\tilde 0\rangle$. Excited states of the dot are formed by means of creation of electrons in the higher level, and/or by creation of holes, i.e.\ by removing electrons, from the lower level. This gives, in addition to $|\tilde 0\rangle$ and $|e\rangle$, $|\bar e\rangle$, $|h\rangle$, $|\bar h\rangle$, the further dot states
\begin{eqnarray*}
&& |e\bar e\rangle , \;\; |h\bar h\rangle , \;\; |eh\rangle \;\; |\bar e\bar h\rangle , \;\; |e\bar h\rangle 
, \;\; |\bar eh\rangle , \\
&& |e\bar eh\rangle , \;\; |e\bar e\bar h\rangle , \;\; |eh\bar h\rangle , \;\; |\bar eh\bar h\rangle , \;\; 
|e\bar eh\bar h\rangle .
\end{eqnarray*}

\begin{figure}[htb]
\centerline{\includegraphics[width=7cm]{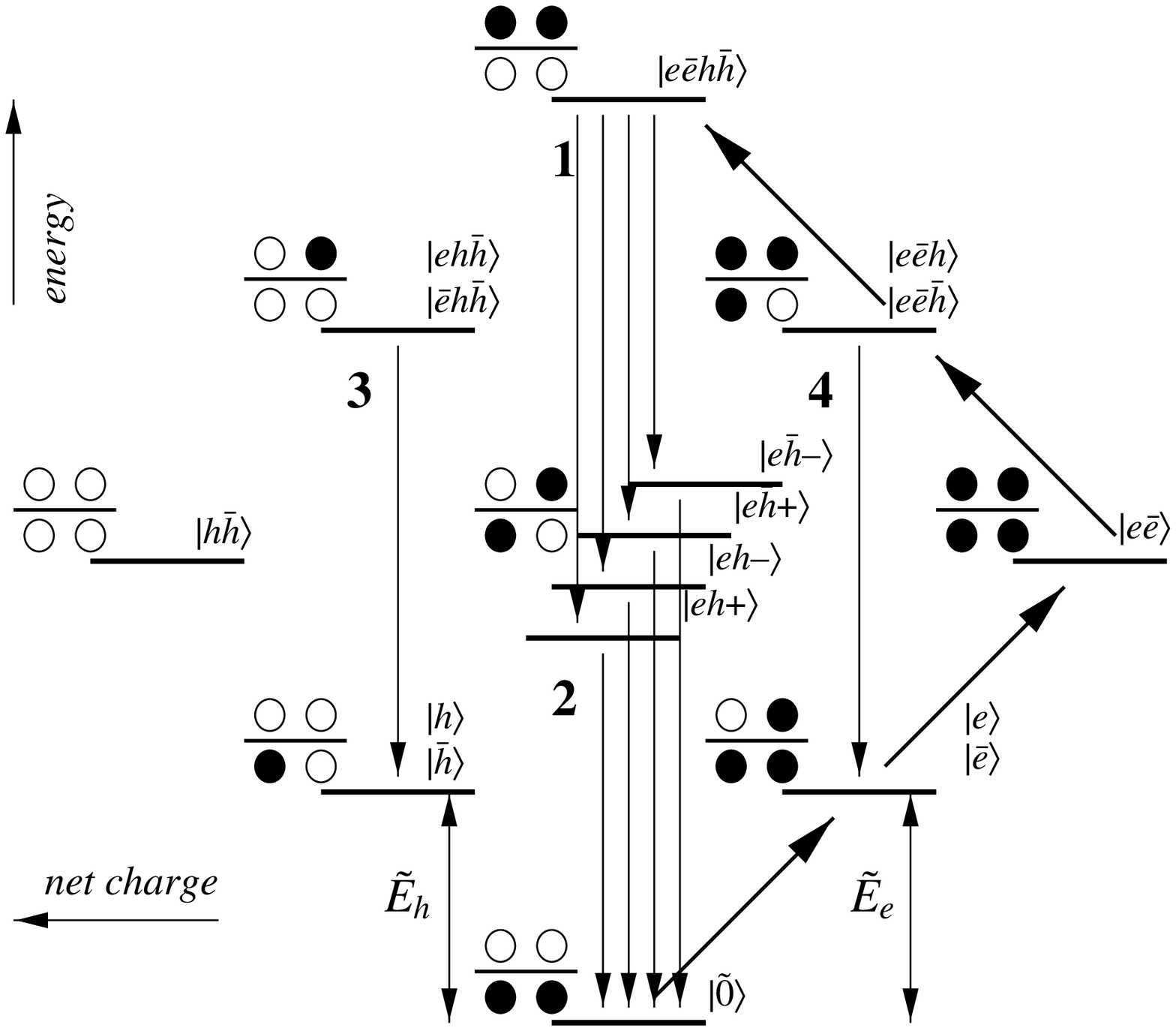}}
\caption{States in a quantum dot model with one (twofold degenerate) electron level and one (twofold degenerate) hole level. The presence of an electron is indicated by a black dot above the line; a hole by an open circle below the horizontal line. Besides the neutral states in the dot, there are states with charge $\pm e$, $\pm 2e$ where one or two excess electrons or holes are present. States with charge $\pm e$ are twofold degenerate due to time-reversal symmetry. The one-electron plus one-hole states are classified according to their even ($+$) or odd ($-$) behavior under time-reversal. Downward arrows indicate photon emission; diagonal arrows indicate a possible path to reach the upper state by subsequent tunneling of two electrons and two holes from the conduction and valence band into the dot.}
\label{Fig2}
\end{figure}

The exciton (one electron plus one hole) states may be split up in energy by the effective interaction between electrons and holes, in the form of second quantization:
\[
\mathbf V = \textstyle\frac{1}{4} \displaystyle\sum_{\alpha\beta\gamma\delta} V_{\alpha\beta\gamma\delta} 
\mathbf a^\dagger_\alpha \mathbf a^\dagger_\beta \mathbf a_\gamma \mathbf a_\delta ,
\]
where the labels $\alpha$, $\beta$, $\gamma$, and $\delta$ stand for $e$, $\bar e$, $h$, or $\bar h$. In first approximation the antisymmetrized matrix elements $V_{\alpha\beta\gamma\delta}$ are those of the (screened) Coulomb interaction, but a more detailed calculation should include many-body effects. However, whatever effects are included, in absence of external magnetic fields the time-reversal symmetry is always conserved. The one-exciton states are therefore either even ($+$) or odd ($-$) under time reversal and therefore given by
\begin{eqnarray}
|eh{+}\rangle &=& (|eh\rangle + |\bar e\bar h\rangle)/\sqrt{2} , \;\; |eh{-}\rangle = 
(|eh\rangle - |\bar e\bar h\rangle)/\sqrt{2} ,
\nonumber \\
|e\bar h{+}\rangle &=& (|e\bar h\rangle - |\bar eh\rangle)/\sqrt{2} , \;\; |e\bar h{-}\rangle  = (|e\bar 
h\rangle + |\bar eh\rangle)/\sqrt{2} .
\label{11}
\end{eqnarray}
For the same reason, the states with charge $\pm 1$ are all twofold degenerate. The time-reversal symmetry implies relations like $V_{eheh}=V_{\bar e\bar h\bar e\bar h}$ and $V_{eh\bar e\bar h}=V_{\bar e\bar heh}$. The energies relative to that of the ground state for the scheme in Fig.\ \ref{Fig2} may than be written in a closed expression as
\begin{eqnarray}
E &=& \tilde E_e n_e + \tilde E_h n_h + \textstyle\frac{1}{2}(n_e-1)n_e V_{e\bar ee\bar e} + \frac{1}{2}(n_h-1)n_h V_{h\bar hh\bar h}
\nonumber \\
&+& (s-\textstyle\frac{1}{2}n_e n_h) (V_{eheh}+V_{e\bar he\bar h}) + \frac{1}{2}s^2(V_{eheh} - V_{e\bar he\bar h}) 
\nonumber \\
&+& \textstyle \frac{1}{2}(1+s)t V_{eh\bar e\bar h} + \frac{1}{2}(1-s)t V_{e\bar h\bar eh} .
\label{1a}
\end{eqnarray}
In this expression the symbols $n_e$, $n_h$ represent the number of electrons in the upper level and the number of holes in the lower level ($n_e,n_h=0,1,2$). For $n_e,n_h=1,1$ we introduced in Eq.\ (\ref{1a}) the notation $t=\pm 1$ for time-even or odd states and the symbol $s$ has the value $s=1$ for $|eh t\rangle$ and $s=-1$ for $|e\bar ht\rangle$ states. For $n_e,n_h\neq 1,1$, we substitute $t=s=0$ in Eq.\ (\ref{1a}). This splitting of the one-exciton multiplet, shown in Fig.\ \ref{Fig2}, is consistent with the phenomenological Hamiltonian used in refs.\ \cite{Poel,Foxon,Bayer}.

\begin{figure}[htb]
\centerline{\includegraphics[width=8cm]{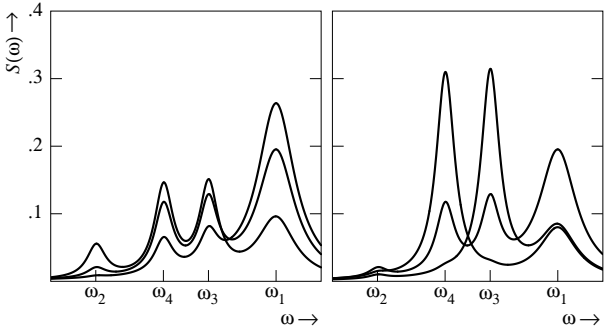}}
\caption{Typical patterns of the photon emission spectrum for a (lens-shaped) quantum dot, corresponding to level scheme \ref{Fig4}a) in the regime of strong tunneling. The plots are based the populations given in paragraph \ref{IIIA} and on the relations (\ref{18}), (\ref{19}) for the photon energies. The relative distances between the latter are arbitrarily chosen. The left plots shows the effect of increasing the bias potential above the resonance condition $eV=\tilde E_e+\tilde E_h$ by $6$, $4$ and $2$ times $k_{\rm B}T$. The right plots show, for $eV=\tilde E_e+\tilde E_h+2k_{\rm B}T$, the effect of changing the gate potential $e\Phi$ by an amount $-k_{\rm B}T$ (curve peaked at $\omega_3$) to $+k_{\rm B}T$ (curve peaked at $\omega_4$) and zero (peaked at $\omega_1$, as in the left figure).}
\label{Fig3}
\end{figure}

If the basis functions (\ref{2}) have no other quantum numbers, corresponding to symmetries of the system, then in principle all the optical transitions indicated in Fig.\ \ref{Fig2} will be present. This means that there is then no dark exciton state. However, not all transition amplitudes in the decay of the biexciton states will be equally large. If symmetries that we consider in the following are fulfilled, some states will be dark. It should be mentioned that experimental information can be obtained by means of polarization measurements in combination with an external magnetic field \cite{Bayer}. According to Eq.\ (\ref{1a}) the sum of the photon energies of the transitions between the charged states is equal to the sum of the energies of the cascade photons in the decay of the (neutral) biexciton state:
\begin{equation}
\omega_1 + \omega_2 = \omega_3 + \omega_4 .
\label{18}
\end{equation}
Because holes are heavier than electrons, their wave function are more confined leading to a stronger repulsion between two holes than that between two electrons, $V_{h\bar hh\bar h}>V_{e\bar ee\bar e}>0$. From this follow with Eq.\ (\ref{1a}) the inequalities
\begin{equation}
\omega_1 - \omega_2 > \omega_3 - \omega_4 > 0 .
\label{19}
\end{equation}
The relations (\ref{18}) and (\ref{19}) are expected to hold under the general condition that the dot is small compared to the bulk exciton size, irrespective of the shape of the dot or crystal structure and may therefore be helpful to analyze the emission spectrum, when other information is lacking. Typical patterns of the emission spectra should then look similar to those plotted in Fig.\ \ref{Fig3}.

In the following paragraph we discuss the qualitatively different schemes that can occur for cylindrically symmetric dots, and consider the implications for generation of entangled photon pairs. Total absence of spatial symmetry and negligible spin-orbit interaction give rise to an unfavourable situation.  For then the twofold degenerate electron and hole sates (\ref{2}) may be written in a simpler form, without $\psi_2$ and $\chi_2$ and with real spatial functions $\psi_1$ and $\chi_1$. In that case antisymmetrization implies that two electrons or holes in the same level form a spin singlet state $S=0$. The one-electron plus one-hole states are then a spin singlet $S=0$, corresponding to $|e\bar h{+}\rangle$ of Eq.\ (\ref{11}) and a spin triplet $S=1$, corresponding to $|eh\rangle$, $|\bar e\bar h\rangle$ and $|e\bar h{-}\rangle$. Such a situation is depicted in Fig.\ \ref{Fig4}d). Since the electric dipole operator does not act on the spin degrees of freedom, the $S=1$ triplet will be dark and the de-exitation of the biexciton state proceeds only via the $S=0$ exciton state. If no other (spatial) symmetries are present, this is only one single state and consequently entanglement of the cascade photons can never occur as that requires two different, energetically indistinguishable paths of the biexciton decay. We therefore conclude that for a possible entanglement of the cascade photons at least spin-orbit interaction, i.e.\ nonzero $\psi_2$ and $\chi_2$ in Eqs.\ (\ref{2}), or some spatial symmetry of the dot potential is required.

\begin{widetext}

\begin{figure}[htb]
\centerline{\includegraphics[width=8cm]{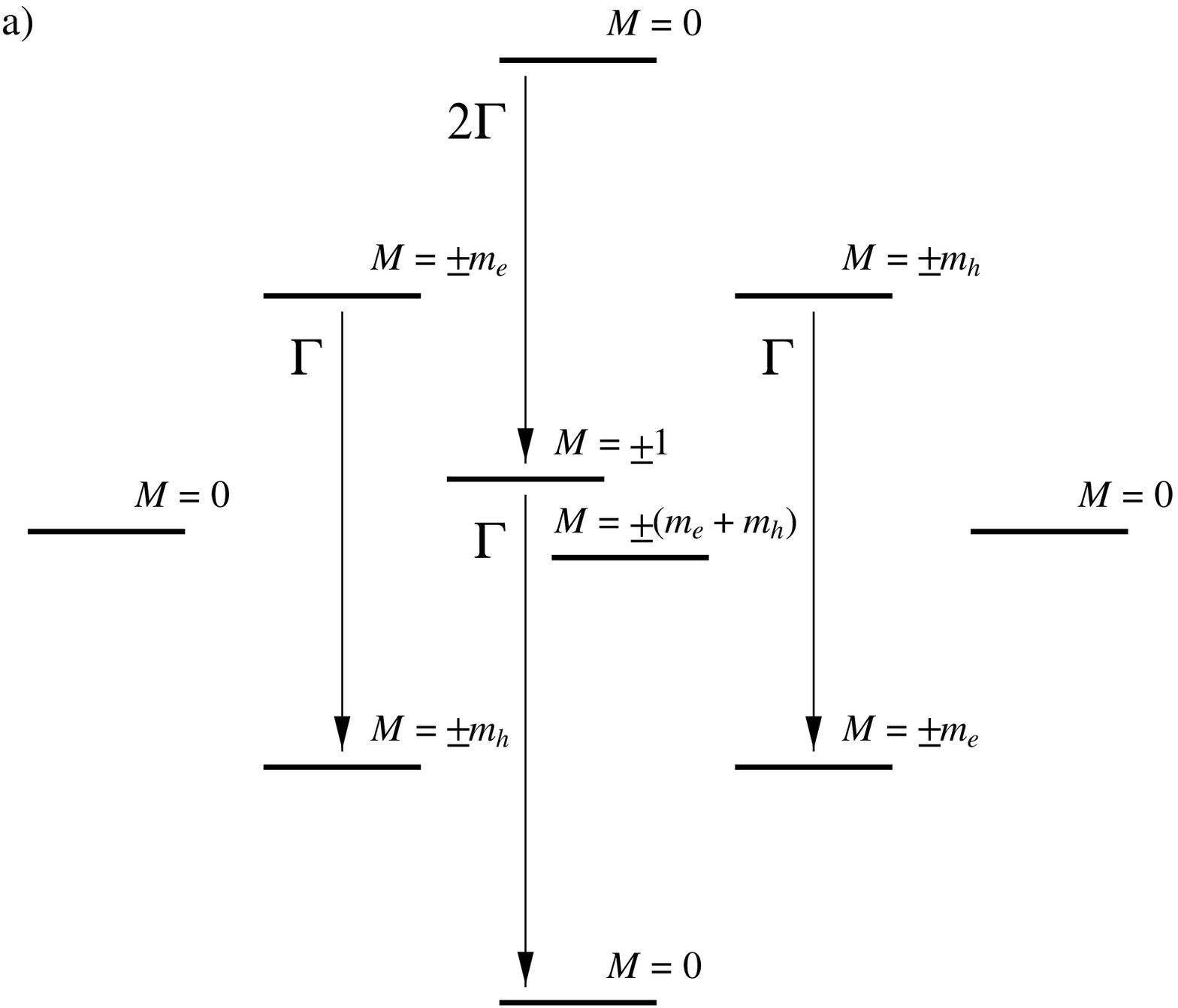}\includegraphics[width=8cm]{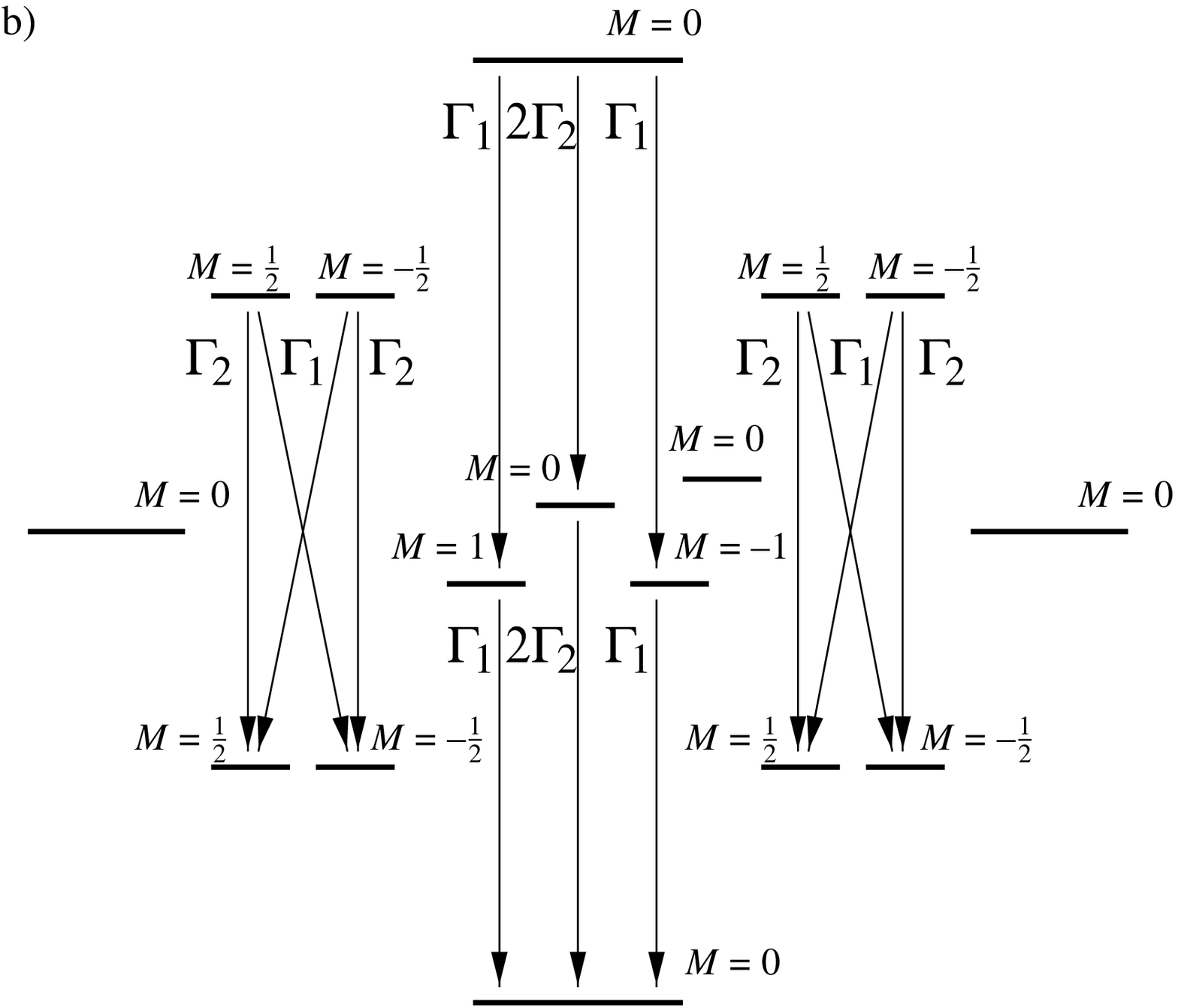}}
\centerline{\includegraphics[width=8cm]{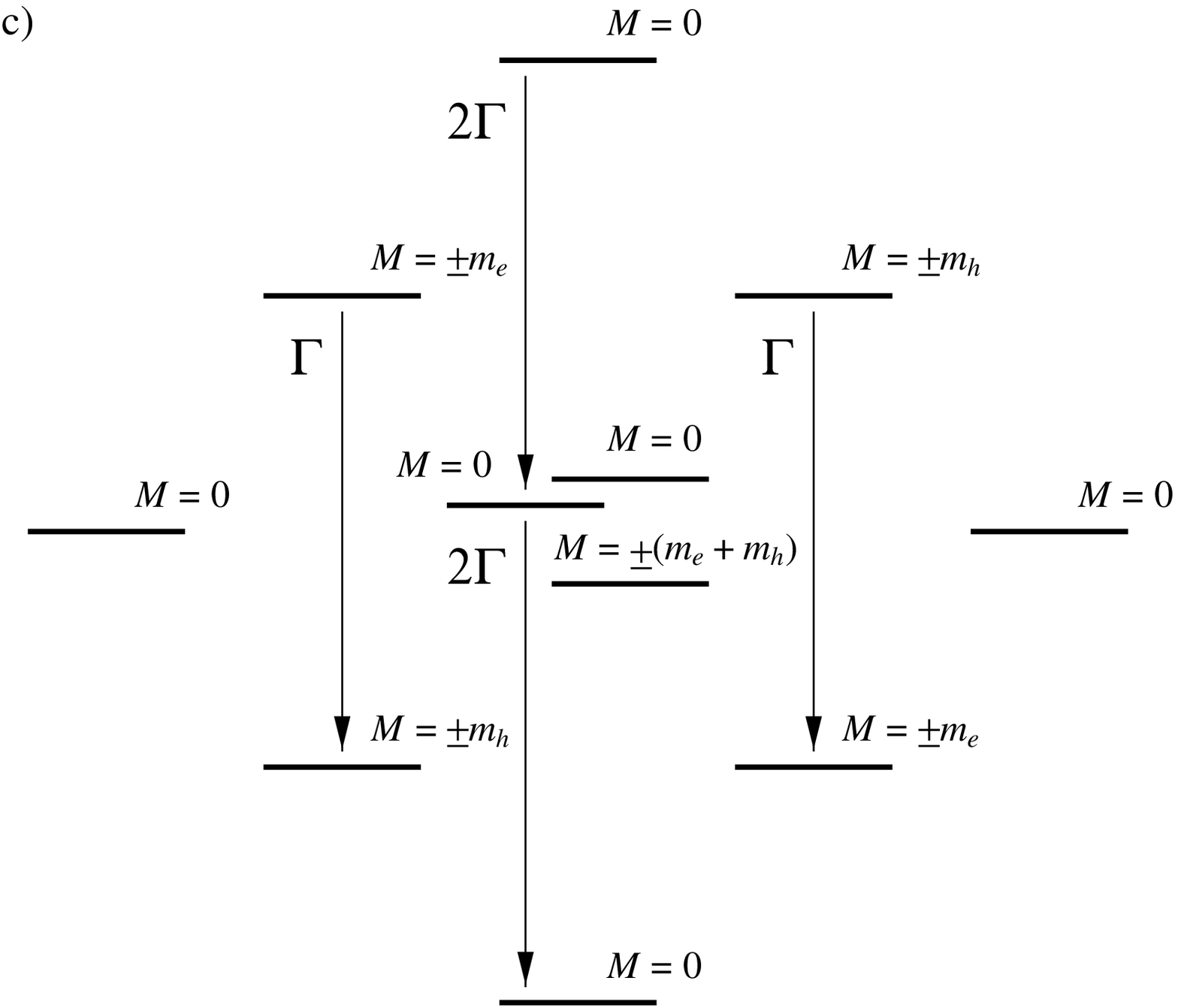}\includegraphics[width=8cm]{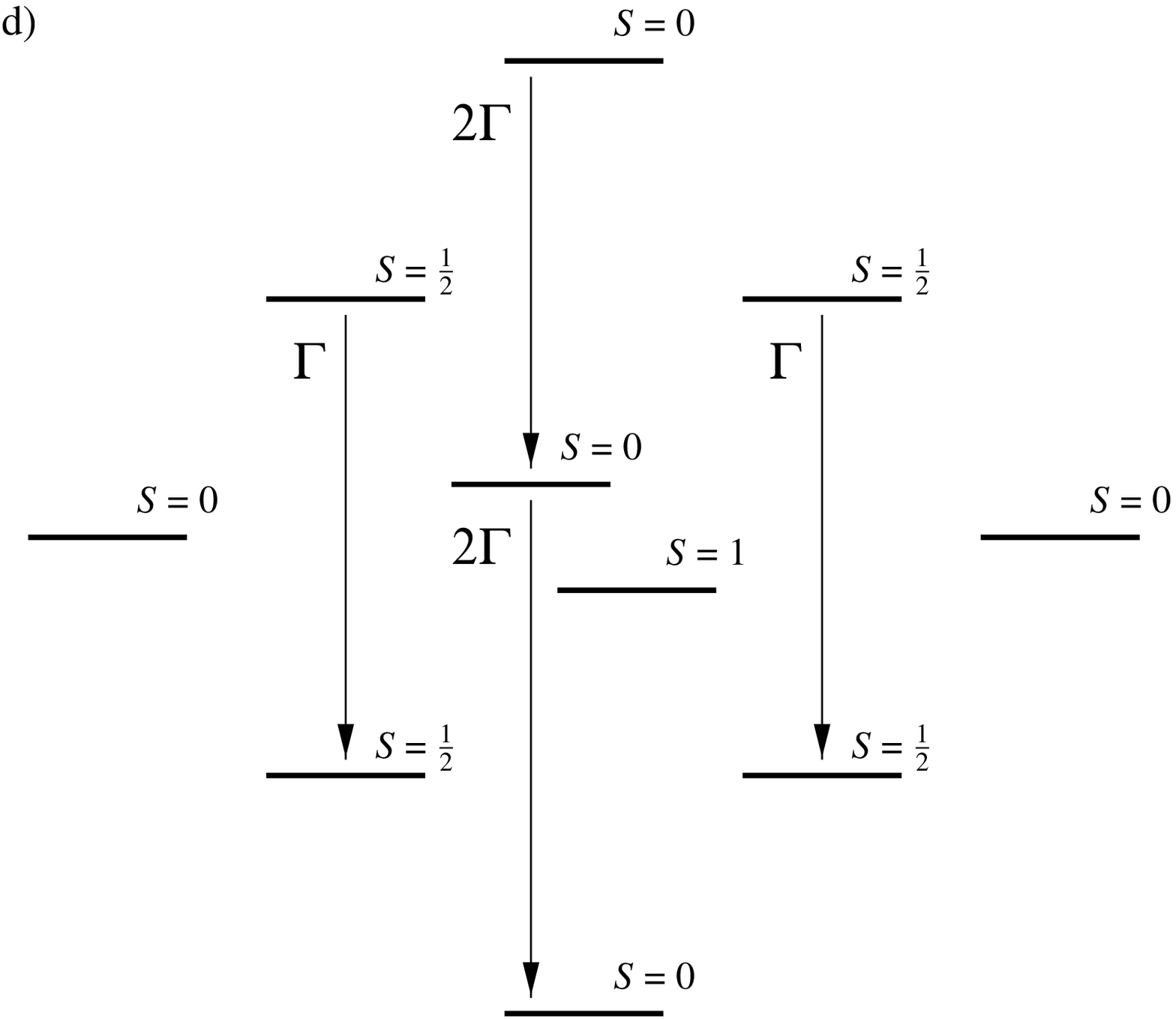}}
\caption{Four different schemes for cylindrical dots. a) In the case $|m_e-m_h|=1$, the exciton multiplet consists of a bright and a dark doublet. b) For the case $m_e=m_h=\frac{1}{2}$, the exciton level is split into a doublet and two singlets. One singlet is a dark state. This results in six optical emission frequencies. c) In case $m_e=m_h>\frac{1}{2}$, the exciton level is split into a doublet and two singlets. Only one of the exciton states (a singlet) is bright. d) Level scheme for systems where spin-orbit coupling can be neglected and total spin $S$ is a good quantum number. If no other symmetries are present, there is only one cascade decay path from the biexciton state. Hence, schemes c) and d) do not produce entangled photons.}
\label{Fig4}
\end{figure}


\begin{figure}[htb]
\centerline{\includegraphics[width=8cm]{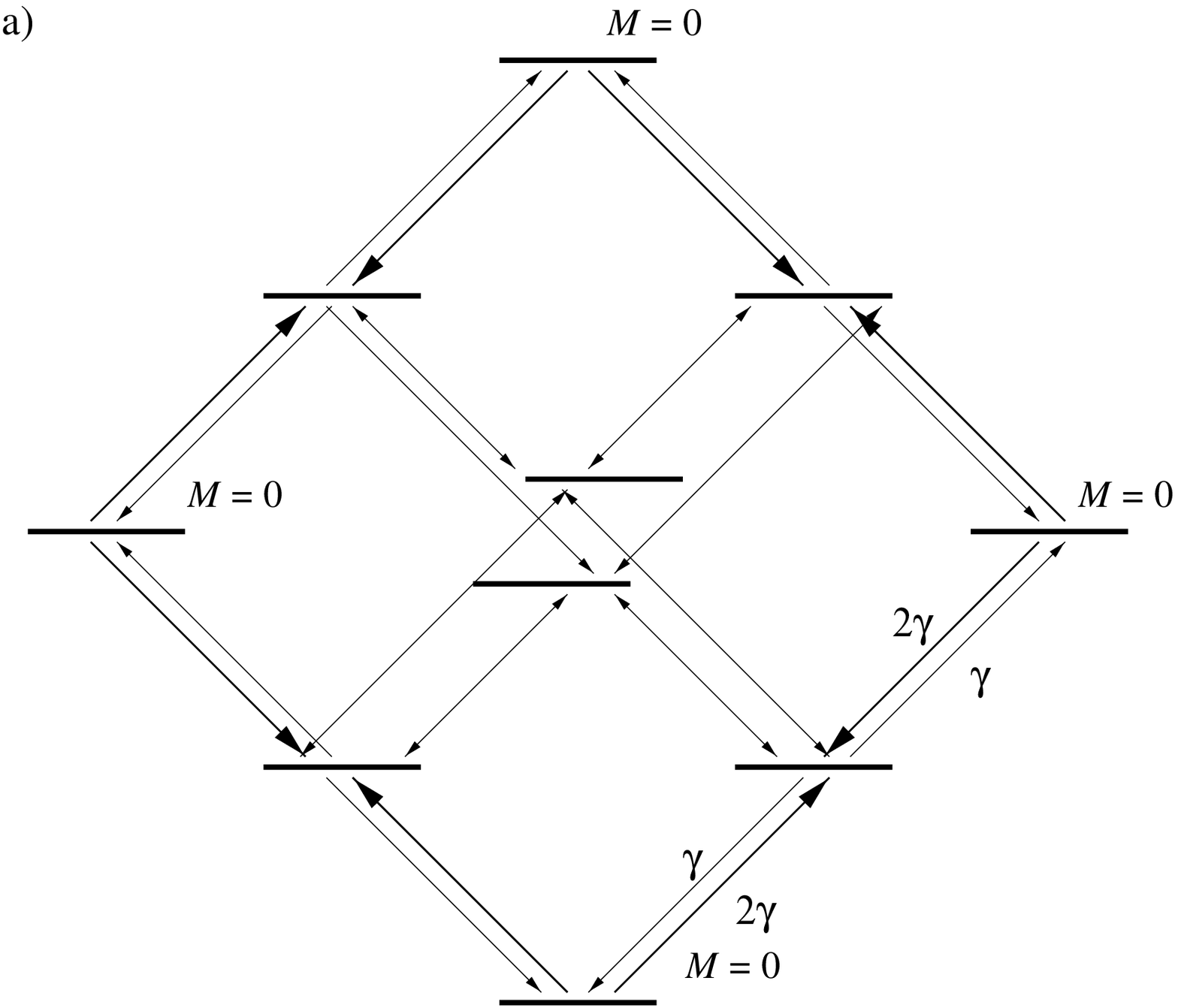}
\includegraphics[width=8cm]{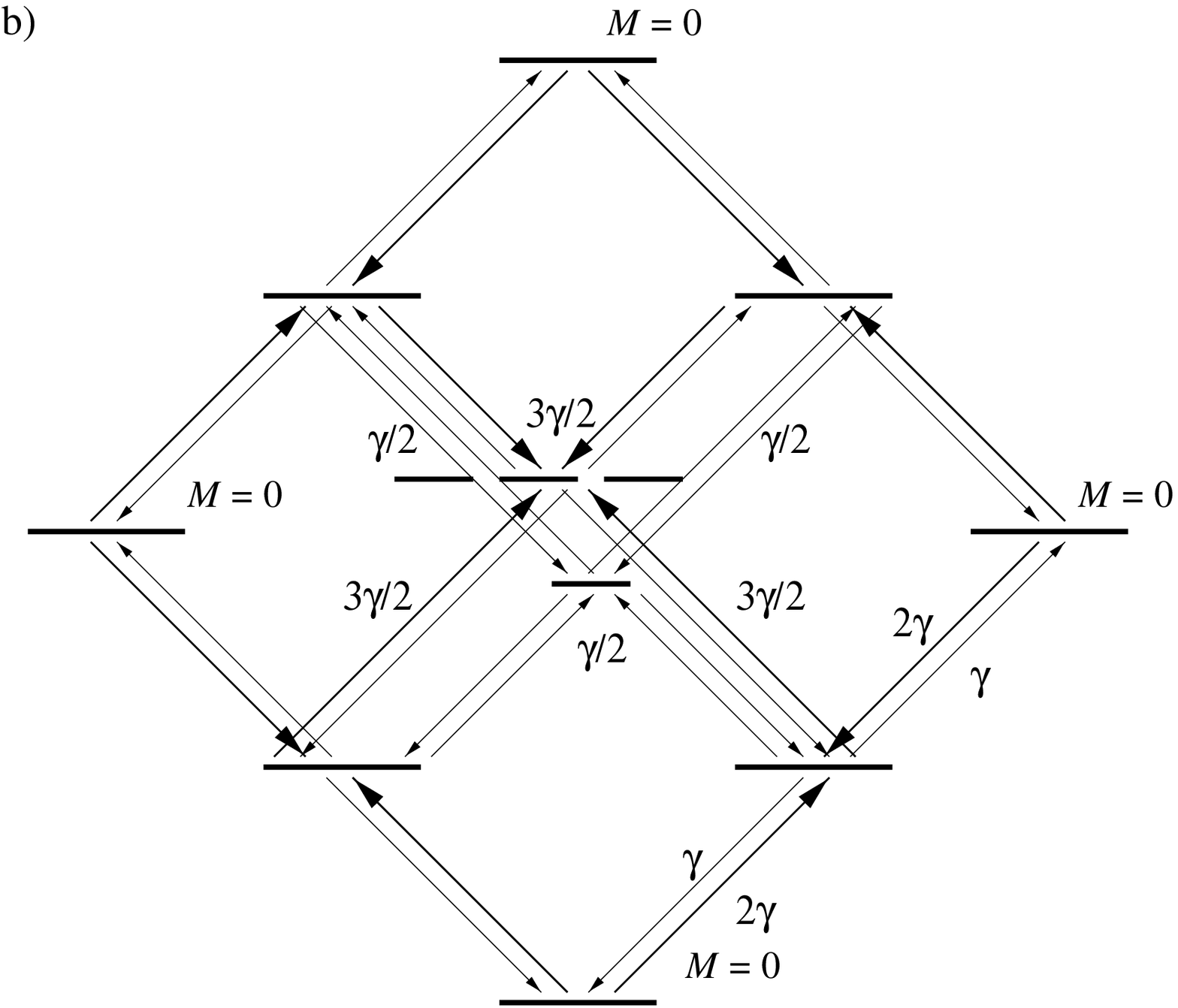}}
\caption{Tunneling rates, for resonant tuning. a) The case $m_e\neq m_h$: the bright and a dark exciton doublets are reached by equal tunneling probabilities. b) The case $m_e=m_h=\frac{1}{2}$: the bright triplet and dark singlet are reached with unequal rates. Unless indicated otherwise, thin arrows have rate $\gamma$, thick arrows have rate $2\gamma$.}
\label{Fig5}
\end{figure}

\end{widetext}

\subsection{Cylindrically Symmetric Dots}

In order to create entanglement in photon pairs in the cascade of Fig.\ \ref{Fig2}, a number of conditions must be satisfied. Firstly the two photons must be identified as coming from transition $1$ and $2$ and must belong to the same cascade. Incoherent tunneling effects between the exciton states must occur at a slow rate $\gamma$ with respect to the lifetime of the exciton level. Secondly, two {\it paths} in the cascade must be indistinguishable, which implies that the splittings $\Delta$ between the intermediate exciton substates should be not greater than the line widths. If no temporal or spatial separation of the photons is possible, one relies on spectral separation, which leads to the conditions $\omega_1-\omega_2\gg\Gamma$, $\Gamma\gg\Delta$, and $\Gamma\gg\gamma$. In the general case with only time-reversal degeneracy of the levels, shown in Fig.\ \ref{Fig2}, the level splittings $\Delta$ and $\omega_1-\omega_2$ will generally be of comparable size, since they are caused by the same effective interaction between the charge carriers. Identification of entangled pairs will also be complicated by the fact that there are four different routes from the biexciton to the ground state. For this reason we consider the case of axially symmetric quantum dots, which are often realized in experiments \cite{Bayer,Kouwenhoven}. As we shall see, this leads to dipole forbidden transitions, i.e.\ dark states, and degeneracy of two exciton states, so that the above conditions for entanglement can be satisfied. 

In an axially symmetric quantum dot, the electron and hole states are characterized by well defined magnetic quantum numbers $\pm m_e$ and $\pm m_h$ respectively. In the most commonly used semiconductor materials the conduction band corresponds to $s\frac{1}{2}$, while the valence band is a $p\frac{3}{2}$ hole band. Hence the first unoccupied level in the dot is an $s\frac{1}{2}$ state, while the highest occupied level will be a $p\frac{3}{2}$ state. For $s\frac{1}{2}$ electrons and $p\frac{3}{2}$ holes, the single-particle states (\ref{3}) are of the form
\begin{eqnarray}
|e\rangle &=& \sum_{\vec b} \sum_{m=-\frac{1}{2}}^{\frac{1}{2}} \int\!\!d\vec r\, |\vec r\rangle \langle\vec r-\vec b|s\textstyle\frac{1}{2},m\rangle \psi_m(\vec b) ,
\nonumber \\
|h\rangle &=& \sum_{\vec b} \sum_{m=-\frac{3}{2}}^{\frac{3}{2}} \int\!\!d\vec r\, |\vec r\rangle \langle\vec r-\vec b|p\textstyle\frac{3}{2},m\rangle \chi_m(\vec b) .
\label{3}
\end{eqnarray}
The summation is over the lattice sites $\vec b$. The state at each lattice site is determined by the slowly varying amplitudes $\psi_m(\vec b)$, or $\chi_m(\vec b)$, and by the localized orbitals $|s\frac{1}{2},m\rangle$, $|p\frac{3}{2},m\rangle$. The latter are the Wannier functions \cite{Ashcroft} which, in the tight-binding approximation, may be replaced by the orbitals for an isolated atom. The spin-components $\psi_m(\vec r)$, $\chi_m(\vec r)$ are called the envelope wave functions \cite{Vahala}. Because these are slowly varying with respect to the lattice, one may replace the argument $\vec b$ with $\vec r$ in the above expressions. One finds that each component of the the envelope wave function is multiplied with a lattice periodic function, which are the Bloch states at the symmetry point $\vec p=\vec 0$. Projection of the states in (\ref{3}) onto the spinor basis gives the components $\psi_1$, $\psi_2$, and $\chi_1$, $\chi_2$ of the general expression (\ref{2}). The electron envelope wave functions $\psi_m(\vec r)$ and the hole envelope wave functions $\chi_m(\vec r)$ are determined by solving an effective Schr\"odinger equation with an added potential $U(\vec{\mathbf r})$ that describes the position-dependent band edge. This leads to the confined dot states. The kinetic energy operator in the equation for the holes is given by the Luttinger Hamiltonian \cite{Haug,Koch,Janssens}
\begin{eqnarray}
&& \mathbf H = \frac{\vec{\mathbf p}^2}{2m^*} - \frac{1}{9\gamma_1 m^*} \sum_{ij=1}^3 [\gamma_3-(\gamma_3-\gamma_2)\delta_{ij}] \mathbf T_{ij} \mathbf J_{ij} ,
\nonumber \\
&& \mathbf T_{ij} = 3\mathbf p_i \mathbf p_j - \delta_{ij} \vec{\mathbf p}^2, \;\;
\mathbf J_{ij} = \frac{3}{2} ( \mathbf j_i \mathbf j_j + \mathbf j_j \mathbf j_i) - \frac{9}{4}\delta_{ij} .
\label{12}
\end{eqnarray}
The Luttinger constants $\gamma_1$, $\gamma_2$ and $\gamma_3$ are dimensionless model parameters. The momentum operator $\vec{\mathbf p}$ may be interpreted as a quantization of the Bloch momentum, because a plane wave envelope function corresponds to a Bloch wave. It is important to realize, however, that the physical electron position and momentum operators act on both the envelope wave functions and the orbitals in the states (\ref{3}).

In the envelope description of localized states in the quantum dot, one may define a total (envelope plus orbital) angular momentum operator as
\[
\vec{\mathbf f} = \vec{\mathbf l} + \vec{\mathbf j} = \vec{\mathbf r}\times\vec{\mathbf p} + \vec{\mathbf j} 
.
\]
Only in the case $\gamma_2=\gamma_3$, the three components of $\vec{\mathbf f}$ commute with $\mathbf H$ \cite{Vahala,Koch}. Because for InAs the two constants are nearly equal, this so called spherical approximation is often made. In the spherical approximation, a cylindrical confinement potential $U$ gives rise a the constant of motion $m_f$; a spherical confinement potential $U$ results in constant $f$ and $m_f$. In realistic calculations for the case of a spherical dot \cite{Koch}, one finds that the lowest state of the exciton as well as the lowest exciton and biexciton states are predominantly composed of an $l=0$ envelope wave function. In that case, the angular momentum of ground states roughly equals that of the orbital functions $f=j$.

If we restrict ourselves to cylindrical dots, the confinement potential $U$ is axially symmetric and the single-particle states have good quantum numbers $m_f=m_e$ and $m_f=m_h$ for the electron and the hole. These single-particle states will be denoted as
\[
|e\rangle = |m_e\rangle , \;\; |\bar e\rangle = |-m_e\rangle , \;\; |h\rangle = |m_h\rangle , 
\;\; |\bar h\rangle = |-m_h\rangle ,
\]
with positive $m_e$, $m_h$. From the sixteen basis states of the level scheme Fig.\ \ref{Fig2}, the states with an even number of electrons and of holes have total magnetic quantum number $M=0$. One also has that $V_{eh\bar e\bar h}$ vanishes. As a consequence the pairs of opposite $M$ in the one-exciton multiplet, like for example $|eh\rangle$ and $|\bar e\bar h\rangle$, are degenerate. There are qualitatively different schemes, shown in Fig.\ \ref{Fig4}. Since electric dipole transitions occur only if $|m_e-m_h|\leq 1$, we distinguish two cases: $|m_e-m_h|=1$, diagram \ref{Fig4}a), and $m_e=m_h$, diagrams \ref{Fig4}b) and \ref{Fig4}c). The case $|m_e-m_h|=1$ is realized in the (lens-shaped) In(Ga)As/(Ar)GaAs quantum dots that have been extensively studied in ref.\ \cite{Bayer}. There it was found that for zero external magnetic field the exciton states with $|M|=m_e+m_h=2$, which are formed by a $m=\frac{3}{2}$ heavy-hole state and a $m=\frac{1}{2}$ electron state, are to good approximation dark and lie below the $|M|=1$ bright exciton states. So diagram \ref{Fig4}a) represents a realistic situation and the dark excitons with $M=\pm(m_e+m_h)$ may even act as intermediate stages in the formation of the biexciton state by tunneling of electrons and holes into the dot. This is also a favorable situation for the creation of entangled photons, since the two exciton states with $|M|=1$ are degenerate, due to time-reversal symmetry, and the decay of the biexciton via the $M=+1$ and via the $M=-1$ exciton state are therefore indistinguishable, as required for entanglement. This only holds true, of course, if there is perfect axial symmetry \cite{Foxon,Forchel}.

We now consider the situations sketched in diagrams \ref{Fig4}b) and \ref{Fig4}c), which represent the cases $m_e=m_h=\frac{1}{2}$ and $m_e=m_h>\frac{1}{2}$ respectively. In diagram \ref{Fig4}b) there is a doublet of bright states, which may allow for entanglement of photons that are polarized in the horizontal plane. Note that the $m_h=\frac{1}{2}$ state is a superposition of $m_l=-1,0,1,2$ states in the $p\frac{3}{2}$ hole level, while the $m_h=\frac{3}{2}$ state is a superposition of $m_l=0,1,2,3$ states. In dots elongated in the $z$ direction, hereafter called `tall' dots, the $m_f=\frac{1}{2}$ is expected to ly below the $m_f=\frac{3}{2}$ state. In lens-shaped dots, the ground state has $m_f=\frac{3}{2}$ instead. The relevant level in a $p\frac{3}{2}$ hole band may therefore consist of the $m_h=\pm\frac{1}{2}$ states for tall cylindrical dots. The scheme of Fig.\ \ref{Fig4}b) would also occur if, due to strain or other effects, the split-off $p\frac{1}{2}$ band provides the hole states. In diagram \ref{Fig4}b) there are two degenerate exciton states, with $M=+1$ and $M=-1$, which are appropriate for the production of entangled photon pairs. Then one of the $M=0$ exciton states is dark, the other is bright. The energy of the $M=0$ states differs in general from that of the $M=1$ states and therefore in total six frequencies appear in the optical spectrum. In diagram \ref{Fig4}c) the exciton states with $M=\pm(m_e+m_h)$ are obviously dark states, but also the $M=0$ time-odd exciton state $|e\bar h{-}\rangle$ is dark. This follows from the time-reversal property of the dipole operator $e\mathbf z$ of the $M=0$ to $M=0$ transition. So in this case there is only one bright exciton state, $|e\bar h{+}\rangle$, and therefore this situation does not allow production of entangled photon pairs in the cascade decay of the biexciton. We conclude therefore that diagram Fig.\ \ref{Fig4}b) represents a possibly favourable case for the production of entangled photons while a situation as depicted in diagram \ref{Fig4}c) is not suitable.

In case of a spherical quantum dot, an exceptional situation may occur if both the electron level and the hole level are states with angular momentum $f_e=f_h=\frac{1}{2}$. The spherical symmetry then leads to a (threefold degenerate) $F=1$ triplet and one dark $F=0$ exciton
\begin{eqnarray*}
|eh\rangle &=& |FM=11\rangle , \;\; |\bar e\bar h\rangle = |FM=1{-1}\rangle , \\
|e\bar h{-}\rangle &=& - |FM=10\rangle , \;\; |e\bar h{+}\rangle = - |FM=00\rangle .
\end{eqnarray*}
In the de-excitation cascade of the biexciton state now three polarizations are possible for the same photon energy, which yield extra options for entanglement. A spherical quantum dot with $f_e$ or $f_h$ larger than $\frac{1}{2}$ results in a system that can be seen as a combination of several systems with $m_e$ and $m_h$ taking all the possible values. The total number of states is $4^{f_e+f_h+1}$ and the levels have a large degree of degeneracy.

For very small dots of size comparable with the lattice constant, the crystal symmetry will be incompatible with spherical symmetry or cylindrical symmetry. One does not expect that the Luttinger Hamiltonian (\ref{12}) can describe this situation. Such small dots will fall in the class of Fig.\ \ref{Fig2}. Another extreme situation may arize for bound states on a single impurity atom in a further homogeneous crystal. Such a system resembles an ionic atom and has spherical symmetry. Our model applies only to the simplest situation: two levels with $j_e=j_h=\frac{1}{2}$ in scheme Fig.\ \ref{Fig4}b), for example with a s$\frac{1}{2}$ electron level and a p$\frac{1}{2}$ hole level.

\section{Statistics and Entanglement}
\subsection{Emission in the Strong Tunneling Limit}
\label{IIIA}

An important factor that determines to what extent entangled photons will be emitted is the ratio of the spontaneous emission rates $\Gamma$ in Fig. \ref{Fig4} and the tunneling rates $\gamma$ of the charge carriers. We now show how $\Gamma$ may be experimentally determined in a situation of fast tunneling. For preparation of the biexciton the bias voltage is increased to a value where the tunneling rate $\gamma$ is much greater than the photon emission rate $\Gamma$, so that the electron and hole tunneling is fast compared to spontaneous emission. We neglect the nonradiative recombination \cite{Takagahara}. In this regime, only thermal fluctuations can de-excite the system \cite{Beenakker}. When thermal energy exceeds the Coulomb shifts, $k_{\rm B}T\gg V_{\alpha\beta\gamma\delta}$, the populations of the single particle states are independent and equal the Fermi-Dirac distribution in the continuum bands:
\begin{equation}
p_e = \frac{1}{\displaystyle 1 + \exp\frac{\tilde E_e-e\Phi-eV/2}{k_{\rm B}T} } , \;\;
p_h = \frac{1}{\displaystyle 1 + \exp\frac{\tilde E_h+e\Phi-eV/2}{k_{\rm B}T} } .
\label{22}
\end{equation}
The decaying levels: the biexciton, the bright exciton, and the two charged excitons, then have respective populations $p_e^2p_h^2$, $2p_e(1-p_e)p_h(1-p_h)$, and $2p_e(1-p_e)p_h^2$, $2p_e^2p_h(1-p_h)$ for a flat dot (the system with $|m_e-m_h|=1$). For a tall dot, or another realization of the $m_e=m_h=\frac{1}{2}$ scheme, the population of the bright exciton is $3p_e(1-p_e)p_h(1-p_h)$ instead. Multiplication of these populations with the decay rate for each of the levels as indicated in Fig.\ \ref{Fig4}a) and b), gives the strength of the emission peaks. Examples of emission spectra in thermal equilibrium for the case of strong tunneling are shown in Fig.\ \ref{Fig3}. The average emission time of a photon as a function of temperature equals $\bar t=1/2\Gamma p_e p_h$ for a flat dot and is $\bar t=1/2(\Gamma_1+\Gamma_2)p_e p_h$ for a tall dot. By measuring this average one can experimentally determine $\Gamma$, respectively $\Gamma_1+\Gamma_2$. Clearly, for the preparation of a pure biexciton state, $p_e$ and $p_h$ must be close to one.

\subsection{Correlated Photon Pairs}

\begin{figure}[htb]
\centerline{\includegraphics[width=8cm]{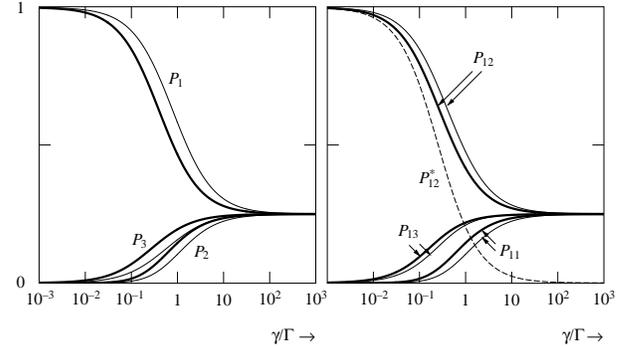}}
\caption{Photon emission probabilities for transitions $1$, $2$ and $3$, as function of the ratio of the carrier tunneling rate $\gamma$ and the photon emission rate $\Gamma$ of the lower transition as given by Eqs.\ (\ref{31})-(\ref{33}). Thick and thin lines correspond to flat and tall dots. For tall dots we adopted $\Gamma_1=\Gamma_2=\Gamma$. Left plots: emission of the first photon after preparation of the biexciton; right plots: emission of the second photon, when the first was emitted on transition $1$. The dotted line is the probability that no tunneling event occurs between transition $1$ and transition $2$.}
\label{Fig6}
\end{figure}

We now suppose that the quantum dot has been prepared in the biexciton state, so that the two-photon cascade can be detected. Any residual tunneling of electrons and holes can result in a tunneling out of the intermediate one-exciton state and lead to emission of a photon from another transition. At the resonance (\ref{4}), the states in the continuum levels are half filled, i.e.\ Eq.\ (\ref{22}) gives $p_e=p_h=\frac{1}{2}$. Then, the unconditional tunneling probability of an electron into or out of the dot is the same and the perturbations are minimal. To obtain an analytical estimate for the relative photon emission probabilities and their correlations, we assume that the electron and hole tunneling have roughly the same rate $\gamma$. For the evaluation of the jump statistics, the system can be described by a classical master equation, since only incoherent transitions occur \cite{Walls,Carmichael}. Since the tunneling rates do not depend on whether an exciton is bright or dark, the populations of the members in the multiplets can simply be added so that the number of rate equations is reduced. The net tunneling between the levels is indicated in Fig.\ \ref{Fig5}, corresponding to the two schemes that can give entanglement in Fig.\ \ref{Fig4}. We calculate the emission probabilities after preparation of the biexciton for each of the four transitions. The probability of a transition between a pair of levels is a matrix element of the inverse of the transition matrix, neglecting the gain terms of photon emission. For flat dots, the scheme of Fig.\ \ref{Fig4}a) and \ref{Fig5}a), we find the expressions
\begin{eqnarray}
P_2 &=& \frac{6\gamma^2}{2\Gamma^2+15\Gamma\gamma+24\gamma^2} , \;\;
P_3 = \frac{2\Gamma\gamma+6\gamma^2}{2\Gamma^2+15\Gamma\gamma+24\gamma^2} ,
\nonumber \\
P_1 &=& 1 - P_2 - 2P_3 , \;\; P_4 = P_3 .
\label{31}
\end{eqnarray}
The probabilities that a photon emission on transition $1$ is followed by each of the other transitions are
\begin{eqnarray}
P_{11} &=& \frac{6\gamma^2}{2\Gamma^2+15\Gamma\gamma+24\gamma^2} , \;\;
P_{13} = \frac{3\Gamma\gamma+6\gamma^2}{2\Gamma^2+15\Gamma\gamma+24\gamma^2} ,
\nonumber \\
P_{12} &=& 1 - P_{11} - 2P_{13} , \;\; P_{14} = P_{13} .
\label{32}
\end{eqnarray}
For tall dots, the scheme of Fig.\ \ref{Fig4}b) and \ref{Fig5}b), the emission probabilities on the four transitions are different, because of the different exciton structure. The probabilities of emission starting from the biexciton or starting from the exciton after transition $1$ are in this case given by
\begin{eqnarray}
P_2 &=& P_{11} = \frac{4\gamma^2}{\Gamma_{\rm t}^2+9\Gamma_{\rm t}\gamma+16\gamma^2} ,
\label{33} \\
P_3 &=& \frac{\Gamma_{\rm t}\gamma+4\gamma^2}{\Gamma_{\rm t}^2+9\Gamma_{\rm t}\gamma+16\gamma^2} , \;\;
P_{13} = \frac{2\Gamma_{\rm t}\gamma+4\gamma^2}{\Gamma_{\rm t}^2+9\Gamma_{\rm t}\gamma+16\gamma^2} ,
\nonumber
\end{eqnarray}
where $\Gamma_{\rm t}=\Gamma_1+\Gamma_2$. The dependence of the probabilities in (\ref{31}), (\ref{32}) and (\ref{33}) on the ratio of the residual tunneling rate $\gamma$ and the photon emission rate $\Gamma$ (see Figs.\ \ref{Fig4} and \ref{Fig5}) is plotted in Fig.\ \ref{Fig6}. It appears that the photon correlation $P_{12}$ is greater than $90\%$ when $\Gamma>10\gamma$ but falls to $25\%$ when the tunneling rate is much faster than the photon decay. Even if the first two photons are on the cascade transition $1$ followed by $2$, this does not yet guarantee entanglement. We calculate the degree of entanglement in the next paragraph.

The resonant tunneling rate $\gamma$, for electrons and holes, may be experimentally determined from the average time between two subsequent photon emissions in the steady state regime. This average time difference is for flat and for tall dots respectively given by the following two expressions:
\begin{equation}
\bar t = \frac{1}{\gamma} + \frac{2}{\Gamma} ,
\;\;
\bar t = \frac{8}{9\gamma} + \frac{2}{\Gamma_{\rm t}} + \frac{2}{9}\, \frac{5\Gamma_{\rm t}+24\gamma}{3\Gamma_{\rm t}^2+28\Gamma_{\rm t}\gamma+48\gamma^2} .
\label{44}
\end{equation}

\subsection{Entangled Photon Pairs}

\begin{figure}[htb]
\centerline{\includegraphics[width=8cm]{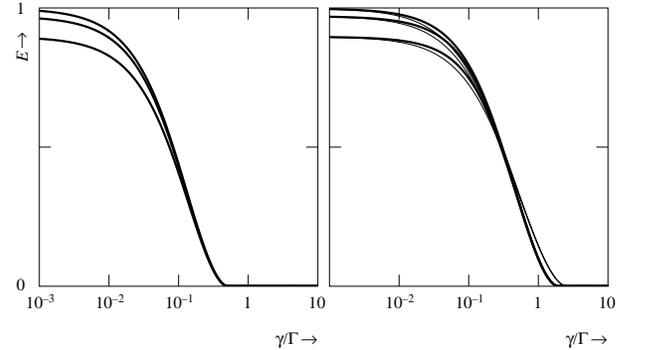}}
\caption{Dependence of the entanglement entropy (\ref{40}) on the ratio $\gamma/\Gamma$, where for tall dots $\Gamma=\Gamma_1$. The upper, middle and lower curves are for an exciton energy splitting of $\Delta=0$, $\Delta=.2\Gamma$, and $\Delta=.4\Gamma$ respectively. The left plot applies when photons from the decay of the charged exciton states ($3$ and $4$ in Fig.\ \ref{Fig2}) are not detected. The right plot applies when these photons are eliminated. Thick lines refer to flat dots, scheme of Fig.\ \ref{Fig4}a), thin lines to tall dots, scheme of Fig.\ \ref{Fig4}b), with axial symmetry.}
\label{Fig7}
\end{figure}

During the switching interval that allows tunneling of two electrons and two holes, the system is prepared in the biexciton state. This is followed by spontaneous emission of one photon and the system makes the transition
\begin{eqnarray*}
|e\bar eh\bar h\rangle &\rightarrow & (|e\bar h{+}\rangle |x\rangle + |e\bar h{-}\rangle |y\rangle)/\sqrt{2} , \;\; ({\rm flat\ dot}) ,
\\
&\rightarrow & (|eh{+}\rangle |x\rangle + |eh{-}\rangle |y\rangle)/\sqrt{2} , \;\; ({\rm tall\ dot}) .
\end{eqnarray*}
Here $|x\rangle$ and $|y\rangle$ are orthogonal linear polarization vectors of the radiation field. We consider here only the case of observation of photons emitted along the $z$ direction. As a result, an entangled state between the dot and the electromagnetic field is formed. In order to have a degenerate doublet of bright one-exciton states in the the one-exciton multiplet, one needs an axially symmetric dot. Any asymmetry gives rise to a splitting of the bright doublet of magnitude $\Delta=2V_{eh\bar e\bar h}$ (flat dot) or $\Delta=2V_{e\bar h\bar e h}$ (tall dot). We consider first the case that the tunneling rate $\gamma$ is small compared to the energy splitting $\Delta$, so that we can neglect tunneling effects. When the system resides in the one-exciton state for a time $t$, the state will evolved into the state
\[
(|e\bar h{+}\rangle|x\rangle e^{-i\Delta t} + |e\bar h{-}\rangle |y\rangle/\sqrt{2} , \;\; {\rm or} \;\;
(|eh{+}\rangle|x\rangle e^{-i\Delta t} + |eh{-}\rangle |y\rangle/\sqrt{2},
\]
for the respective cases of flat and tall dots. The probability for a waiting time $t$ between the two photon emissions in the cascade equals $\Gamma e^{-\Gamma t}$, where $\Gamma=\Gamma_1$ for tall dots. Therefore, the two photon density operator is given by the average
\begin{eqnarray}
\boldsymbol\rho &=& \textstyle\frac{1}{2} \displaystyle \int_0^\infty\!\!\!\!dt\, \Gamma e^{-\Gamma t} \big(|xx\rangle e^{-i\Delta t}+|yy\rangle\big)\big(\langle xx| e^{i\Delta t}+\langle yy|\big)
\label{50} \\
&=& \displaystyle\frac{1}{2+2i\Delta/\Gamma}|xx\rangle\langle yy| + {\rm c.c.} + 
\textstyle\frac{1}{2} |xx\rangle\langle xx| + \frac{1}{2} |yy\rangle\langle yy| .
\nonumber
\end{eqnarray}
This expression shows that dephasing destroys the off-diagonal matrix element, and thereby the entanglement. The correlation between the polarizations $|x\rangle$ and $|y\rangle$ remain perfect. We now include the residual tunneling from the one-exciton doublet to the four charged dot states, which occurs at a total rate of $4\gamma$. When the system (eventually) returns to the bright exciton doublet, the entanglement between the dot and the field is destroyed and also the correlation has disappeared. The probability that the second photon follows the first without a tunneling event is $P^*_{12}=\Gamma/(\Gamma+4\gamma)$, while the second photon is independent of the first with probability $1-P^*_{12}$. If we include this effect, the field density operator has a fraction $1-P^*_{12}$ that is fully mixed state, and a fraction $P^*_{12}$ that is the average with waiting times $t$ with probabilities $\Gamma e^{-(\Gamma+4\gamma)t}$. This gives the two-photon density operator
\begin{eqnarray*}
\boldsymbol\rho &=& \textstyle\frac{1}{2}P \Big( 
\displaystyle\frac{1}{1+i\Delta/(\Gamma+4\gamma)}|xx\rangle\langle yy| + {\rm c.c.} + 
|xx\rangle\langle xx| + |yy\rangle\langle yy| \Big)
\\
&+& \textstyle\frac{1}{4}(1-P) \big(|x\rangle\langle x|+|y\rangle\langle y|\big) \big(|x\rangle\langle 
x|+|y\rangle\langle y|\big)
\end{eqnarray*}
and $P=P^*_{12}$. It is clear from this expression that $P$ is a measure of the polarization correlation. The evaluation of the entanglement entropy $E$ of such a mixed state was described in ref.\ \cite{Wootters}. After a short calculation we obtain $E$ in terms of the concurrence $C$:
\begin{eqnarray}
E &=& -x\log_2 x -(1-x)\log_2(1-x) , \;\;
x = \textstyle\frac{1}{2} + \frac{1}{2}\sqrt{1-C^2} ,
\nonumber \\
C &=& \frac{P}{|1+i\Delta/(\Gamma+4\gamma)|}  - \frac{1-P}{2} .
\label{40}
\end{eqnarray}
When expression (\ref{40}) becomes negative, $C$ and $E$ are defined to be zero. For pure states the concurrence $C$ gives the visibility in two-photon interferometry \cite{Teich}. One finds that for any value of $\Delta$, entanglement is totally destroyed when $2\gamma\geq\Gamma$. This can be seen in the left graph of Fig.\ \ref{Fig7}. As expected, the improvement becomes considerable when the tunneling rate $\gamma$ is comparable with the photon emission rate $\Gamma$.

One may improve on the efficiency of entangled pairs by detecting the photons on transitions $3$ and $4$. Then one can ignore events where the tunneling from the one-exciton level leads to a photon on transition $3$ or $4$ and only count pairs of photons on the cascade $1$ to $2$. The entanglement entropy is still given by Eq.\ (\ref{40}), but with $P$ the conditional probability of an immediate pair $1$, $2$ under the assumption that the second photon is of transition $2$. Hence, this is $P=P^*_{12}/P_{12}$ for flat dots and, provided one detects the horizontally polarized photons, one must put $P=3P^*_{12}/(2P_{12}+P^*_{12})$ for tall dots. Here $P_{12}$ is given in Eq.\ (\ref{32}) and Eq.\ (\ref{33}) the respective two types of level schemes. The result is plotted in the right graph of Fig.\ \ref{Fig7}. As expected, the improvement becomes considerable when the tunneling rate $\gamma$ is comparable with the photon emission rate $\Gamma$.

\subsection{Quantum Dot in an Optical Microcavity}

\begin{figure}[htb]
\centerline{\includegraphics[width=4cm]{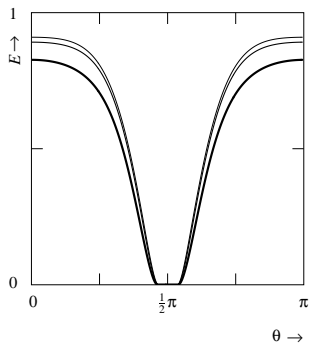}}
\caption{Dependence of the entanglement entropy on the relative orientation, given by the angles $\theta$ and $\phi$, of the cavity with respect to the dot. The upper, middle and lower curves are for $\Delta=.1\Gamma$, $\Delta=.2\Gamma$, $\Delta=.4\Gamma$ respectively. The plots are for a constant $\phi=\pi/4$ with a tunneling rate chosen at $\gamma=.01\Gamma$.}
\label{Fig8}
\end{figure}

Application of an optical microcavity (resonant with the lower transition $2$ of the cascade), such as dielectric Bragg mirrors or a photonic crystal, increases the decay rate $\Gamma$ and therefore is another means to enhance the entanglement entropy. The cavity may also enhance the relative emission in a specific spatial direction. If, however, the cavity does not have two degenerate polarization modes in the $xy$ plane, the level scheme of a flat dot Fig.\ \ref{Fig4}a) is perturbed and the entanglement is corrupted. For example, let us assume that the symmetry axis of the cavity is misalinged (with respect to the $z$ axis of the dot) in the direction $\hat z\cos\theta+(\hat x\cos\phi+\hat y\sin\phi)\sin\theta$. This implies that the dipole transitions corresponding to the polarizations $|u\rangle=|x\rangle\cos\phi+|y\rangle\sin\phi$ and $|v\rangle=|y\rangle\cos\phi-|x\rangle\sin\phi$ have modified coupling constants so that the decay rates in this basis are $\Gamma\cos^2\theta$ and $\Gamma$. With quantum trajectory techniques \cite{Carmichael,Visser} we obtain for the density operator
\begin{eqnarray*}
\boldsymbol\rho &=& \int_0^\infty\!\!\!\!dt\, \sqrt{\mathbf G} e^{-i\mathbf H t-\mathbf G t/2} \boldsymbol\sigma  e^{i\mathbf H t-\mathbf G t/2-4\gamma t} \sqrt{\mathbf G} +  (1-P) {\bf 1} ;
\\
\boldsymbol\sigma &=& \textstyle\frac{1}{2} \big(|xx\rangle+|yy\rangle\big)\big(\langle xx|+\langle yy|\big) ,
\\
{\bf 1} &=& \textstyle\frac{1}{4} \big(|x\rangle\langle x|+|y\rangle\langle y|\big) \big(|x\rangle\langle 
x|+|y\rangle\langle y|\big) ,
\end{eqnarray*}
which is a straitforward generalization of expression (\ref{50}). The Hamiltonian $\mathbf H$ and the decay operator $\mathbf G$ act on the state of photon $2$ only:
\[
\mathbf H = |x\rangle\langle x|\Delta , \;\; \mathbf G = |u\rangle\langle u|\Gamma\cos^2\theta + |v\rangle\langle v|\Gamma ,
\]
and $P$ is determined from normalization. Plots of the entanglement entropy for various misalignment angles are shown in Fig.\ \ref{Fig8}. Due to the complicated resonance structure of a cavity, the transitions $1$, $3$ and $4$ will generally have different decay rates. These are preferrably smaller than the modified spontaneous emission rate $\Gamma$ of transition $2$, so that $P_{11}$, $P_{13}$ and $P_{14}$ are small and $P_{12}$ is nearly unity.

\section{Conclusions}

We considered realizations of a two-photon turnstile based on small quantum dots. In the regime of tight confinement, the single-particle states are well separated and the Coulomb interaction can be treated perturbatively. We showed how this results in a closed level scheme with sixteen base states. The system seems ideal for generation of entangled photons on the cascade from the biexciton via the excitonic multiplet to the ground state. The biexciton can be prepared without Coulomb blockade so that low temperatures are not needed. For a cylindrically (but not spherically) symmetric dot, different combinations of the magnetic electron and hole quantum numbers $m_e$, $m_h$ give rise to the four different level schemes depicted in Figure \ref{Fig4}. Selection rules for optical transitions imply that only in the first two cases, with $m_e-m_h=\pm 1$ or $m_e=m_h=\frac{1}{2}$, a degenerate intermediate level occurs in the cascade, which is a requirement for entanglement. Quantum dots in (In)GaAs/(Al)GaAs with a flat cylindrical shape have $m_e=\frac{1}{2}$, $m_h=\frac{3}{2}$ electron and hole ground states, while tall dots that are elongated along the symmetry axis, have $m_e=m_h=\frac{1}{2}$ due to restricted orbital angular momentum. Therefore, both level schemes, Fig.\ \ref{Fig4}a) and \ref{Fig4}b), can be realized experimentally.

The polarization correlation and entanglement of formation in the photon pair may be corrupted by the following two effects; firstly here will be a minimal residual tunneling rate $4\gamma$ into and out of the intermediate one exciton level, which can effectively flip the spin of the exciton. Secondly, the Coulomb interaction gives rise to an exchange splitting of the exciton multiplet in dots without perfect axial symmetry, which causes different polarization states to dephase. The residual tunneling rate $\gamma$ may be obtained from the emission statistics of pairs different from the cascade $1$ followed by $2$ as given by Eq.\ (\ref{44}). The polarization correlation is found to be as much as $75\%$ if $\gamma\leq 0.1\Gamma$ and drops only about $20\%$ if $\gamma\approx\Gamma$. The entanglement entropy of the two photons is still roughly $80\%$ for $\gamma\leq 0.1\Gamma$, provided that photons emitted from charged states of the dot can be eliminated. Otherwise is is roughly halved. The entanglement is rather insensitive to an energy splitting $\Delta$ of the (bright) exciton substates, as long as $\Delta\leq 0.4\Gamma$.

Application of an optical microcavity that is resonant with the lower transition of the cascade, leads to increased $\Gamma$ and thereby enhances the entanglement of the emitted cascade photons. Misalignment of the cavity axis with respect to the symmetry axis of the dot does not substantially decrease the entanglement, as long as the mismatch is less than $\pi/2$.

\acknowledgements

This work is part of the research program of the `Stichting voor Fundamenteel Onderzoek der Materie' (FOM), which is financially supported by the `Nederlandse Organisatie voor Wetenschappelijk Onderzoek' (NWO).


\end{document}